\documentclass[11pt,a4paper]{article}
\pdfoutput=1
\usepackage{jinstpub}
\usepackage{siunitx}

\usepackage{graphicx}
\graphicspath{{figures/}}

\usepackage{multicol}
\usepackage{multirow}
\usepackage{subcaption}
\usepackage{import}

\usepackage{graphicx}
\usepackage{placeins}
\usepackage{comment}
\title{Silicon Vertex Detector of the Belle II Experiment}

\author[a,b,1]{S.~Mondal\note{Corresponding author.},}%\affiliation{Dipartimento di Fisica, Universit\acute{a} di Pisa, I-56127 Pisa, Italy}%\affiliation{INFN Sezione di Pisa, I-56127 Pisa, Italy} % Pisa
\author[c]{K.~Adamczyk,}%\affiliation{H. Niewodniczanski Institute of Nuclear Physics, Krakow 31-342, Poland} % Krakow
\author[d]{L.~Aggarwal,}%\affiliation{Panjab University, Chandigarh 160014, India} % PU
\author[e]{H.~Aihara,}%\affiliation{Department of Physics, University of Tokyo, Tokyo 113-0033, Japan} % Tokyo
\author[f]{T.~Aziz,}%\affiliation{Tata Institute of Fundamental Research, Mumbai 400005, India} % Tata
\author[c]{S.~Bacher,}%\affiliation{H. Niewodniczanski Institute of Nuclear Physics, Krakow 31-342, Poland} % Krakow
\author[g]{S.~Bahinipati,}%\affiliation{Indian Institute of Technology Bhubaneswar, Bhubaneswar 752050, India} % Bhubaneswar
\author[a,b]{G.~Batignani,}%\affiliation{Dipartimento di Fisica, Universit\acute{a} di Pisa, I-56127 Pisa, Italy}%\affiliation{INFN Sezione di Pisa, I-56127 Pisa, Italy} % Pisa
\author[h]{J.~Baudot,} %\affiliation{IPHC, UMR 7178, Universit$\acute{e}$ de Strasbourg, CNRS, 67037 Strasbourg, France} %Strabourg
\author[i]{P.~K.~Behera,}%\affiliation{Indian Institute of Technology Madras, Chennai 600036, India} % Chennai
\author[a,b]{S.~Bettarini,}%\affiliation{Dipartimento di Fisica, Universit\acute{a} di Pisa, I-56127 Pisa, Italy}%\affiliation{INFN Sezione di Pisa, I-56127 Pisa, Italy} % Pisa
\author[j]{T.~Bilka,}%\affiliation{Faculty of Mathematics and Physics, Charles University, 121 16 Prague, Czech Republic} % Prague
\author[c]{A.~Bozek,}%\affiliation{H. Niewodniczanski Institute of Nuclear Physics, Krakow 31-342, Poland} % Krakow
\author[k]{F.~Buchsteiner,}%\affiliation{Institute of High Energy Physics, Austrian Academy of Sciences, 1050 Vienna, Austria} % Vienna
\author[a,b]{G.~Casarosa,}%\affiliation{Dipartimento di Fisica, Universit\acute{a} di Pisa, I-56127 Pisa, Italy}%\affiliation{INFN Sezione di Pisa, I-56127 Pisa, Italy} % Pisa
\author[b]{L.~Corona,}%\affiliation{INFN Sezione di Pisa, I-56127 Pisa, Italy} % Pisa
\author[l]{S.~B.~Das,}%\affiliation{Malaviya National Institute of Technology, Jaipur 302017, India} %MNIT
\author[h]{G.~Dujany,} %\affiliation{IPHC, UMR 7178, Universit$\acute{e}$ de Strasbourg, CNRS, 67037 Strasbourg, France} %Strabourg
\author[h]{C.~Finck,} %\affiliation{IPHC, UMR 7178, Universit$\acute{e}$ de Strasbourg, CNRS, 67037 Strasbourg, France} %Strabourg
\author[a,b]{F.~Forti,}%\affiliation{Dipartimento di Fisica, Universit\acute{a} di Pisa, I-56127 Pisa, Italy}%\affiliation{INFN Sezione di Pisa, I-56127 Pisa, Italy} % Pisa
\author[k]{M.~Friedl,}%\affiliation{Institute of High Energy Physics, Austrian Academy of Sciences, 1050 Vienna, Austria} % Vienna
\author[m,n]{A.~Gabrielli,}%\affiliation{Dipartimento di Fisica, Universit\acute{a} di Trieste, I-34127 Trieste, Italy}%\affiliation{INFN Sezione di Trieste, I-34127 Trieste, Italy} % Trieste
\author[n]{B.~Gobbo,}%\affiliation{INFN Sezione di Trieste, I-34127 Trieste, Italy} % Trieste
\author[f]{S.~Halder,}%\affiliation{Tata Institute of Fundamental Research, Mumbai 400005, India} % Tata
\author[o,p]{K.~Hara,}%\affiliation{High Energy Accelerator Research Organization (KEK), Tsukuba 305-0801, Japan} %\affiliation{The Graduate University for Advanced Studies (SOKENDAI), Hayama 240-0193, Japan}  KEK
\author[f]{S.~Hazra,}%\affiliation{Tata Institute of Fundamental Research, Mumbai 400005, India} % Tata
\author[q]{T.~Higuchi,}%\affiliation{Kavli Institute for the Physics and Mathematics of the Universe, University of Tokyo, Kashiwa 277-8583, Japan} % IPMU
\author[k]{C.~Irmler,}%\affiliation{Institute of High Energy Physics, Austrian Academy of Sciences, 1050 Vienna, Austria} % Vienna
\author[o,p]{A.~Ishikawa,}%\affiliation{High Energy Accelerator Research Organization (KEK), Tsukuba 305-0801, Japan} %\affiliation{The Graduate University for Advanced Studies (SOKENDAI), Hayama 240-0193, Japan}  KEK
\author[n]{Y.~Jin,}%\affiliation{Dipartimento di Fisica, Universit\acute{a} di Trieste, I-34127 Trieste, Italy}%\affiliation{INFN Sezione di Trieste, I-34127 Trieste, Italy} % Trieste
\author[c]{M.~Kaleta,}%\affiliation{H. Niewodniczanski Institute of Nuclear Physics, Krakow 31-342, Poland} % Krakow
\author[k]{A.~B.~Kaliyar,}% Vienna
\author[j]{J.~Kandra,}%\affiliation{Faculty of Mathematics and Physics, Charles University, 121 16 Prague, Czech Republic} % Prague
\author[q]{K.~H.~Kang,}%\affiliation{Kavli Institute for the Physics and Mathematics of the Universe, University of Tokyo, Kashiwa 277-8583, Japan} % IPMU
\author[j]{P.~Kody\v{s},}%\affiliation{Faculty of Mathematics and Physics, Charles University, 121 16 Prague, Czech Republic} % Prague
\author[o]{T.~Kohriki,}%\affiliation{High Energy Accelerator Research Organization (KEK), Tsukuba 305-0801, Japan} % KEK
\author[r]{R.~Kumar,}%\affiliation{Punjab Agricultural University, Ludhiana 141004, India} % PAU
\author[l]{K.~Lalwani,}%\affiliation{Malaviya National Institute of Technology, Jaipur 302017, India} %MNIT
\author[t]{K.~Lautenbach,}%\affiliation{Aix Marseille Universit$\acute{e}$ , CNRS/IN2P3, CPPM, 13288 Marseille, France}
\author[t]{R.~Leboucher,}%\affiliation{Aix Marseille Universit$\acute{e}$ , CNRS/IN2P3, CPPM, 13288 Marseille, France}
\author[s]{S.~C.~Lee,}%\affiliation{Department of Physics, Kyungpook National University, Daegu 41566, Korea} % Kyungpook
\author[i]{J.~Libby,}%\affiliation{Indian Institute of Technology Madras, Chennai 600036, India} % Chennai
\author[h]{L.~Martel,} %\affiliation{IPHC, UMR 7178, Universit$\acute{e}$ de Strasbourg, CNRS, 67037 Strasbourg, France} %Strabourg
\author[a,b]{L.~Massaccesi,}%\affiliation{Dipartimento di Fisica, Universit\acute{a} di Pisa, I-56127 Pisa, Italy}%\affiliation{INFN Sezione di Pisa, I-56127 Pisa, Italy} % Pisa
\author[f]{G.~B.~Mohanty,}%\affiliation{Tata Institute of Fundamental Research, Mumbai 400005, India} % Tata
\author[o,p]{K.~R.~Nakamura,}%\affiliation{High Energy Accelerator Research Organization (KEK), Tsukuba 305-0801, Japan} %\affiliation{The Graduate University for Advanced Studies (SOKENDAI), Hayama 240-0193, Japan}  KEK
\author[c]{Z.~Natkaniec,}%\affiliation{H. Niewodniczanski Institute of Nuclear Physics, Krakow 31-342, Poland} % Krakow
\author[e]{Y.~Onuki,}%\affiliation{Department of Physics, University of Tokyo, Tokyo 113-0033, Japan} % Tokyo
\author[q]{F.~Otani,}%\affiliation{Kavli Institute for the Physics and Mathematics of the Universe, University of Tokyo, Kashiwa 277-8583, Japan} % IPMU
\author[a,b]{A.~Paladino$^{\rm A,}$,}%\affiliation{Dipartimento di Fisica, Universit\acute{a} di Pisa, I-56127 Pisa, Italy}%\affiliation{INFN Sezione di Pisa, I-56127 Pisa, Italy} % Pisa
\author[a,b]{E.~Paoloni,}%\affiliation{Dipartimento di Fisica, Universit\acute{a} di Pisa, I-56127 Pisa, Italy}%\affiliation{INFN Sezione di Pisa, I-56127 Pisa, Italy} % Pisa
\author[s]{H.~Park,}%\affiliation{Department of Physics, Kyungpook National University, Daegu 41566, Korea} % Kyungpook
\author[t]{L.~Polat,}%\affiliation{Aix Marseille Universit$\acute{e}$ , CNRS/IN2P3, CPPM, 13288 Marseille, France}
\author[f]{K.~K.~Rao,}%\affiliation{Tata Institute of Fundamental Research, Mumbai 400005, India} % Tata
\author[h]{I.~Ripp-Baudot,} %\affiliation{IPHC, UMR 7178, Universit$\acute{e}$ de Strasbourg, CNRS, 67037 Strasbourg, France} %Strabourg
\author[a,b]{G.~Rizzo,}%\affiliation{Dipartimento di Fisica, Universit\acute{a} di Pisa, I-56127 Pisa, Italy}%\affiliation{INFN Sezione di Pisa, I-56127 Pisa, Italy} % Pisa
\author[o]{Y.~Sato,}%\affiliation{High Energy Accelerator Research Organization (KEK), Tsukuba 305-0801, Japan} %\affiliation{The Graduate University for Advanced Studies (SOKENDAI), Hayama 240-0193, Japan}  KEK
\author[k]{C.~Schwanda,}%\affiliation{Institute of High Energy Physics, Austrian Academy of Sciences, 1050 Vienna, Austria} % Vienna
\author[t]{J.~Serrano,}%\affiliation{Aix Marseille Universit$\acute{e}$ , CNRS/IN2P3, CPPM, 13288 Marseille, France}
\author[q]{T.~Shimasaki,}%\affiliation{Kavli Institute for the Physics and Mathematics of the Universe, University of Tokyo, Kashiwa 277-8583, Japan} % IPMU
\author[o]{J.~Suzuki,}%\affiliation{High Energy Accelerator Research Organization (KEK), Tsukuba 305-0801, Japan} % KEK
\author[o,p]{S.~Tanaka,}%\affiliation{High Energy Accelerator Research Organization (KEK), Tsukuba 305-0801, Japan} %\affiliation{The Graduate University for Advanced Studies (SOKENDAI), Hayama 240-0193, Japan}  KEK
\author[e]{H.~Tanigawa,}%\affiliation{Department of Physics, The University of Tokyo, Tokyo 113-0033, Japan} % Tokyo
\author[a,b]{F.~Tenchini,}%\affiliation{Dipartimento di Fisica, Universit\acute{a} di Pisa, I-56127 Pisa, Italy}%\affiliation{INFN Sezione di Pisa, I-56127 Pisa, Italy} % Pisa
\author[k]{R.~Thalmeier,}%\affiliation{Institute of High Energy Physics, Austrian Academy of Sciences, 1050 Vienna, Austria} % Vienna
\author[f]{R.~Tiwary,}%\affiliation{Tata Institute of Fundamental Research, Mumbai 400005, India} % Tata
\author[o]{T.~Tsuboyama,}%\affiliation{High Energy Accelerator Research Organization (KEK), Tsukuba 305-0801, Japan} % KEK
\author[e]{Y.~Uematsu,}%\affiliation{Department of Physics, The University of Tokyo,113-0033, Japan} % Tokyo 
\author[m,n]{L.~Vitale,}%\affiliation{Dipartimento di Fisica, Universit\acute{a} di Trieste, I-34127 Trieste, Italy}%\affiliation{INFN Sezione di Trieste, I-34127 Trieste, Italy} % Trieste
\author[e]{Z.~Wang,}%\affiliation{Department of Physics, The University of Tokyo,113-0033, Japan} % Tokyo 
\author[u]{J.~Webb,}%\affiliation{School of Physics, University of Melbourne, Melbourne, Victoria 3010, Australia} % Melbourne % KEK Satellite
\author[n]{O.~Werbycka,}%\affiliation{INFN Sezione di Trieste, I-34127 Trieste, Italy} % Trieste
\author[c]{J.~Wiechczynski,}%\affiliation{H. Niewodniczanski Institute of Nuclear Physics, Krakow 31-342, Poland} % Krakow
\author[k]{H.~Yin,}%\affiliation{Institute of High Energy Physics, Austrian Academy of Sciences, 1050 Vienna, Austria} % Vienna
\author[t]{and L.~Zani$^{\rm B,}$}%\affiliation{Aix Marseille Universit$\acute{e}$ , CNRS/IN2P3, CPPM, 13288 Marseille, France} %
\author[]{\\ \vspace{1 mm} (Belle-II SVD Collaboration)}

\affiliation[a]{Dipartimento di Fisica, Universit\`{a} di Pisa, I-56127 Pisa, Italy, $^A$presently at INFN Sezione di Bologna, I-40127 Bologna, Italy}
\affiliation[b]{INFN Sezione di Pisa, I-56127 Pisa, Italy}
\affiliation[c]{H. Niewodniczanski Institute of Nuclear Physics, Krakow 31-342, Poland}
\affiliation[d]{Panjab University, Chandigarh 160014, India} %PU
\affiliation[e]{Department of Physics, University of Tokyo, Tokyo 113-0033, Japan}
\affiliation[f]{Tata Institute of Fundamental Research, Mumbai 400005, India}
\affiliation[g]{Indian Institute of Technology Bhubaneswar, Bhubaneswar 752050, India}
\affiliation[h]{IPHC, UMR 7178, Universit$\acute{e}$ de Strasbourg, CNRS, 67037 Strasbourg, France} %Strabourg
\affiliation[i]{Indian Institute of Technology Madras, Chennai 600036, India}
\affiliation[j]{Faculty of Mathematics and Physics, Charles University, 121 16 Prague, Czech Republic}
\affiliation[k]{Institute of High Energy Physics, Austrian Academy of Sciences, 1050 Vienna, Austria}
\affiliation[l]{Malaviya National Institute of Technology Jaipur, Jaipur 302017, India} %MNIT
\affiliation[m]{Dipartimento di Fisica, Universit\`{a} di Trieste, I-34127 Trieste, Italy}
\affiliation[n]{INFN Sezione di Trieste, I-34127 Trieste, Italy}
\affiliation[o]{High Energy Accelerator Research Organization (KEK), Tsukuba 305-0801, Japan}
\affiliation[p]{The Graduate University for Advanced Studies (SOKENDAI), Hayama 240-0193, Japan} 
\affiliation[q]{Kavli Institute for the Physics and Mathematics of the Universe, University of Tokyo, Kashiwa 277-8583, Japan}
\affiliation[r]{Punjab Agricultural University, Ludhiana 141004, India} % PAU
\affiliation[s]{Department of Physics, Kyungpook National University, Daegu 41566, Korea}
\affiliation[t]{Aix Marseille Universit$\acute{e}$ , CNRS/IN2P3, CPPM, 13288 Marseille, France, $^B$presently at INFN Sezione di Roma Tre, I-00185 Roma, Italy}
\affiliation[u]{School of Physics, University of Melbourne, Melbourne, Victoria 3010, Australia}

\emailAdd{suryamondal@gmail.com}


\abstract{
 
The silicon vertex detector (SVD) is installed at the heart of the Belle II experiment, taking data at the high-luminosity $B$-Factory SuperKEKB since 2019. The SVD is a four-layer double-sided strip detector with tracking and particle-identification capabilities. In this paper, we report on the performance of the reconstruction of SVD hits. The detector has shown a stable and above-99\% hit efficiency, with a large signal-to-noise in all sensors since the beginning of data taking. Cluster position and time resolution have been measured with 2020 and 2022 data and show excellent performance and stability. In particular, the cluster-position resolution is between 7 and \SI{12}{\micro\meter} for the small-pitch sensors, in reasonable agreement with the expectations, while the cluster time resolution is measured to be below \SI{3}{\ns}. The effect of radiation damage is visible, but not affecting the performance.
  As the luminosity increases, higher machine backgrounds are expected and the excellent hit-time
  information in SVD can be exploited for background rejection. In particular, we have recently developed a novel procedure to select hits by grouping them event-by-event based on their time. This new procedure allows a significant reduction of the fake rate, while preserving the tracking efficiency, and it has therefore replaced the previous cut-based procedure.  
  We have developed a method that uses
  the SVD hits to estimate the track time (previously unavailable) and the collision time. It has a similar precision to the estimate
  based on the drift chamber readout but its execution time is three orders of
  magnitude smaller, allowing a faster online reconstruction that is crucial in a high luminosity
  regime. The track time is a powerful information that allows, together with the aforementioned grouping selection, to raise the occupancy limit above that expected at nominal luminosity, leaving room for a safety factor.
  Finally, in June 2022 the data taking of the Belle II experiment was stopped to install a new two-layer DEPFET detector (PXD) and upgrade
  components of the accelerator. The whole silicon tracker (PXD+SVD) has been extracted from Belle II, the new PXD installed, the detector closed and commissioned. We briefly describe the SVD results of this upgrade.}

\proceeding{16$^{\text{th}}$ Topical Seminar on Innovative Particle and Radiation Detectors (IPRD23)\\
  25--29 September 2023\\
  Siena, Italy}

\begin{document}

\maketitle
\flushbottom

\section{Introduction}

The Belle II experiment \cite{abe2010belle} is a particle physics experiment at the intensity frontier, collecting data at the high-luminosity $B$-Factory
SuperKEKB \cite{ohnishi2013accelerator} in Tsukuba, Japan. The main objective of Belle~II is the  search for physics beyond
the Standard Model through high-precision measurements on extensive datasets of mainly $B$ mesons,
$\tau$ leptons and charm hadrons.

SuperKEKB is an asymmetric electron-positron collider
operated at the centre-of-mass energy of the $\Upsilon (4S)$ resonance. Positron and electron beams are accelerated at an energy of \SI{4}{GeV} and \SI{7}{GeV}, respectively, leading to a boosted center-of-mass system  with $\beta\gamma=0.28$. % 
SuperKEKB has achieved the world record instantaneous peak luminosity of \SI{4.7 e 34}{cm^{-2}s^{-1}} in June
2022; its target peak luminosity is \SI{6 e 35}{cm^{-1}s^{-1}}. Belle II has collected \SI{424}{fb^{-1}} of events in the first years of data taking, while the goal is to collect \SI{50}{ab^{-1}} of data in the next decade.

\begin{figure}
  \centering
  \includegraphics[width=0.65\linewidth]{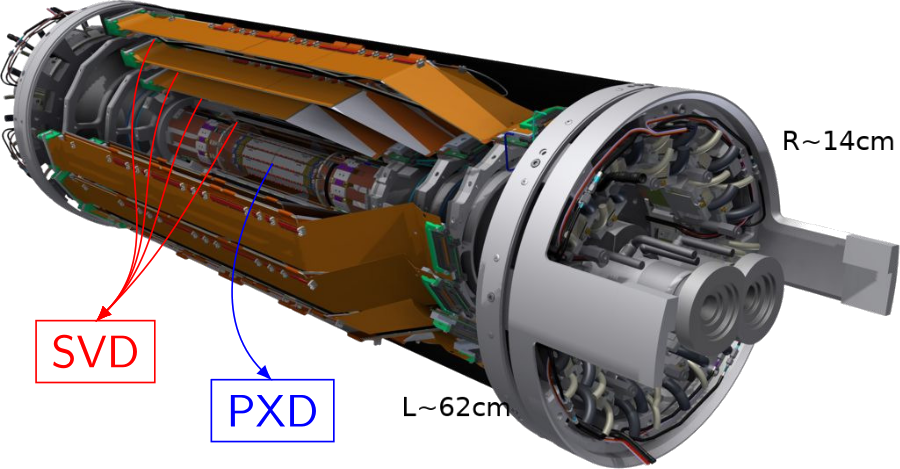}
  \caption{Vertex Detector of Belle II.} \label{fig:svd-3d-sketch}
\end{figure}

The Belle~II detector is designed to provide similar or better performance than its predecessor Belle
in an environment made much harsher by the significantly increased beam background. Belle~II features a superior tracking capability that improves the efficiency of low transverse-momentum tracks and the resolution on the track impact parameters by a factor 2, both thanks to the new Vertex Detector (VXD) shown in Figure~\ref{fig:svd-3d-sketch}.
The VXD also features particle identifications capabilities, through the measurement of the energy loss. This feature is  critical for the low momentum tracks, which do not reach the farther detectors. The VXD consists of two innermost layers of
pixel detector (PXD) made of DEPFET sensors \cite{kemmer1987new}, and four outer layers of double-sided silicon strip detector
(SVD) \cite{adamczyk2022design}.
The main features of the SVD include standalone tracking for low-momentum tracks,
extrapolation of the tracks towards the PXD, precise vertexing of $K_S^0$ and $\Lambda^0$, and particle identification based on $\mathrm{d}E/\mathrm{d}x$ measurement.

\section{Belle II Silicon Vertex Detector}

\begin{figure}
  \centering
  \includegraphics[width=1.0\linewidth]{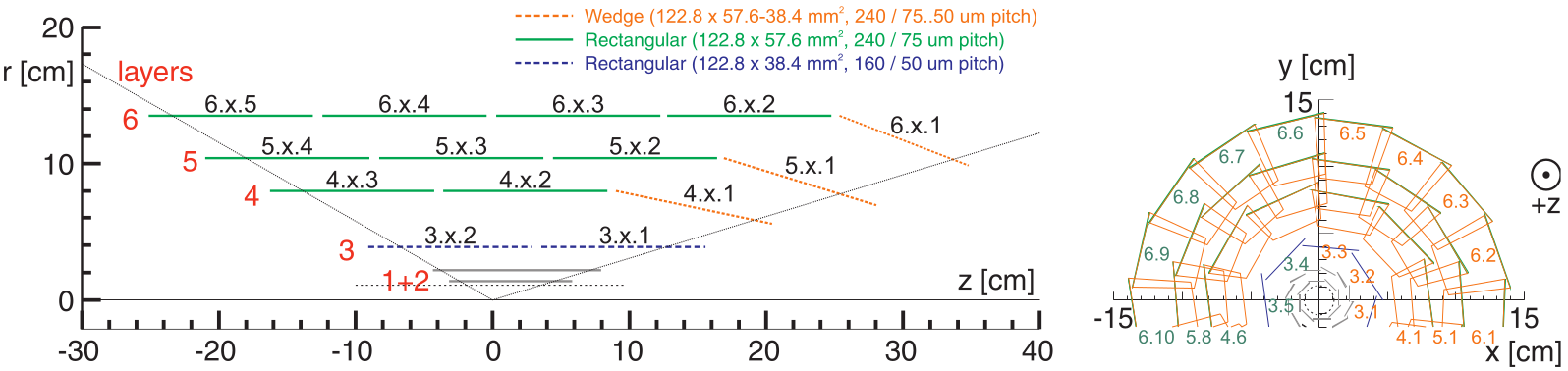}
  \caption{Positions of SVD layers and sensors.} \label{fig:svdhalfsketch}
\end{figure}

The sensitive detectors of the SVD are double sided silicon strip detectors (DSSD),
organised in 4 layers, namely layer 3 (the innermost), 4, 5 and 6 (outermost) and composed respectively of
7, 10, 12 and 16 electrically and
mechanically independent ladders with 2, 3, 4 and 5 sensors each. 
A graphical representation of the placement of the sensors is given
in Figure\,\ref{fig:svdhalfsketch}.
The 172 sensors are covering a sensitive area of \SI{1.2}{m^2} with \num{224e3} readout channels.
The DSSDs provide two-dimensional spatial position information, as well as the collected charge and time of the hits on the two sensor sides. Each sensor is made of an n-type bulk detector with orthogonal strips on the two sides formed with acceptor (P type) or donor (N type) implants. 
The readout strips are AC coupled and there is a floating strip between two adjacent readout strips to improve the determination of the hit position. 
The strips on the acceptor (donor) side, namely u/P (v/N), being parallel (transverse) to the beam
axis,  measure the $r\phi$ ($z$) coordinates of the hits. 

The layer-3 is equipped with small rectangular sensors 
with 768 readout strips on both sides and a readout pitch of 50 (160) \SI{}{\um} on u/P (v/N) side.
The forward sensors in all other layers are slanted and wedge shaped, 
with 768 (512) readout strips and 50-75 (240) \SI{}{\um} readout pitch on u/P (v/N) side.
The sensors in the barrel region are rectangular with 768 (512) readout strips and 75 (240) \SI{}{\um} readout pitch on the u/P (v/N) side.
Sensor thickness is \SI{320}{\um} for small and large sensors and \SI{300}{\um} for slanted sensors.
The full depletion voltages of the sensors is in the range of 20-\SI{60}{V},  with the operating bias
voltage of \SI{100}{V}.

The APV25 \cite{french2001design} chips, with 128 input channels,
are the front-end readout ASICs for the DSSDs. The chip has a fast pulse shaping time of
\SI{50}{\ns}. Being radiation hard up to \SI{100}{Mrad} these chips meet the requirement of the
innermost layer. The APV25 operates in multi-peak mode at a clock frequency of \SI{31.8}{\MHz}
which is $1/8$ of the SuperKEKB bunch-crossing frequency.
Currently, six consecutive samples are read out upon the arrival of the level-1 trigger to reconstruct
the height and time of signal pulses. For higher luminosity runs, a 3/6-mixed operation mode, where three or six samples are acquired depending on the level-1 trigger quality,
is developed for faster readout and smaller data size. 

The SVD adopts the chip-on-sensor ORIGAMI design \cite{adamczyk2022design}, where all the APV25 chips, reading the two sides of the DSSD,  are installed 
on the same side of the sensor. 
The signals from the sensor strips on the two sides are connected to the APV25 chips using flex circuits,  called pitch
adapters. A folded ORIGAMI-like shape is used for the pitch adapters that connect the strip on the other side of the sensor to the side where the APV25 chips are located.  This design allows  cooling of all the readout chips located on the same sensor side using only one cooling pipe and consequently to reduce
 the material budget. The averaged material budget per layer of SVD is {0.7}{\%} of a radiation length.

\section{SVD Operation and Reconstruction Performance}

The SVD was installed in Belle II in November 2018 and it has been acquiring data since March 2019. The operation has been smooth and reliable and the performance of SVD has been consistent with expectations.
The total fraction of masked strips is 
less than {1}{\%}, and they are mainly due to the initial defects caused in the sensor production or ladder assembly, and the hit efficiency exceeds {99}{\%} for all sensors and it is stable with time.
The radiation damage to the
hardware (in terms of increase in strip noise and sensor leakage current) is observed to be at the expected level and has no impact on the reconstruction performance.

\begin{figure}[!htb] 
  \centering
    \includegraphics[width=0.45\linewidth]{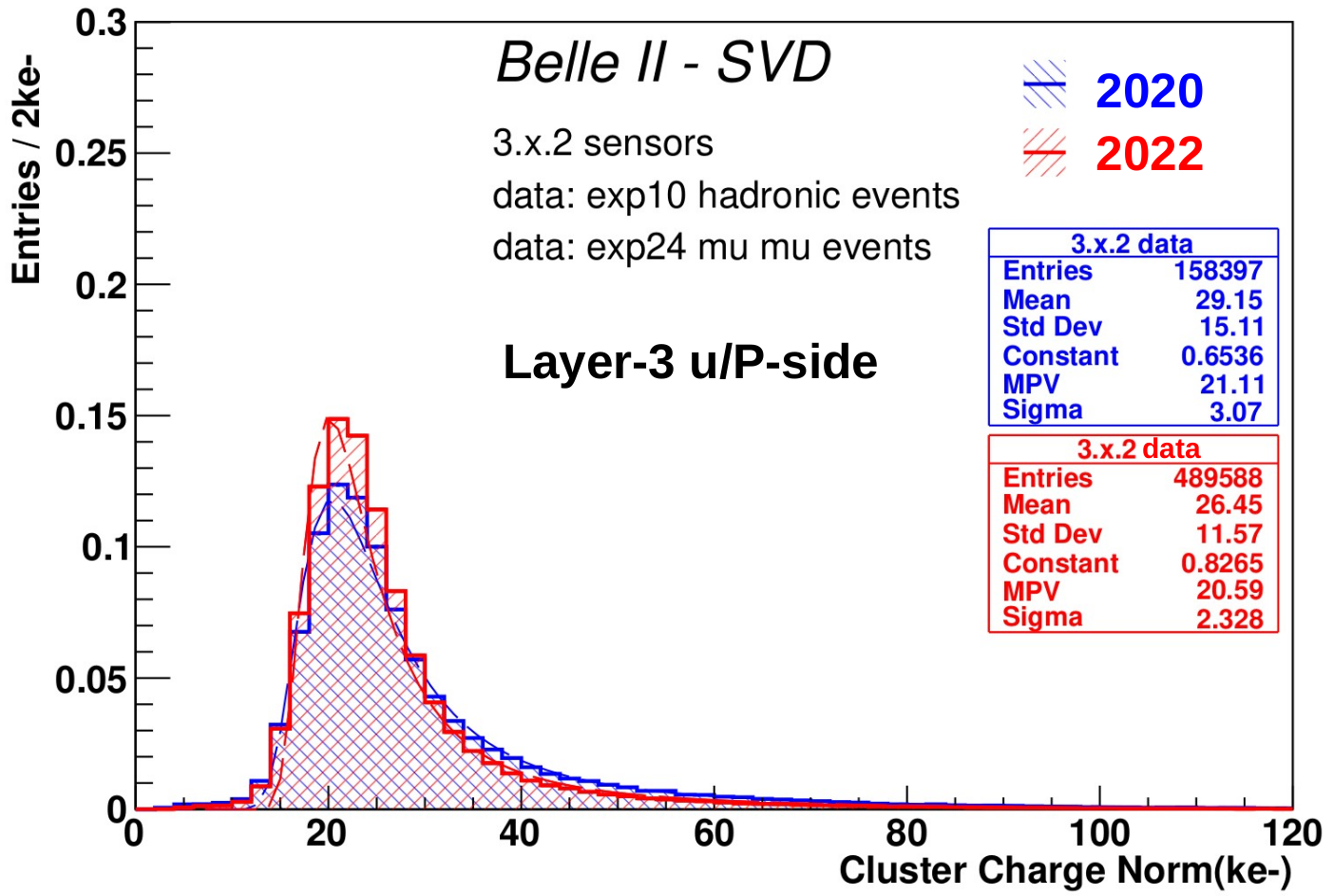}
    \includegraphics[width=0.45\linewidth]{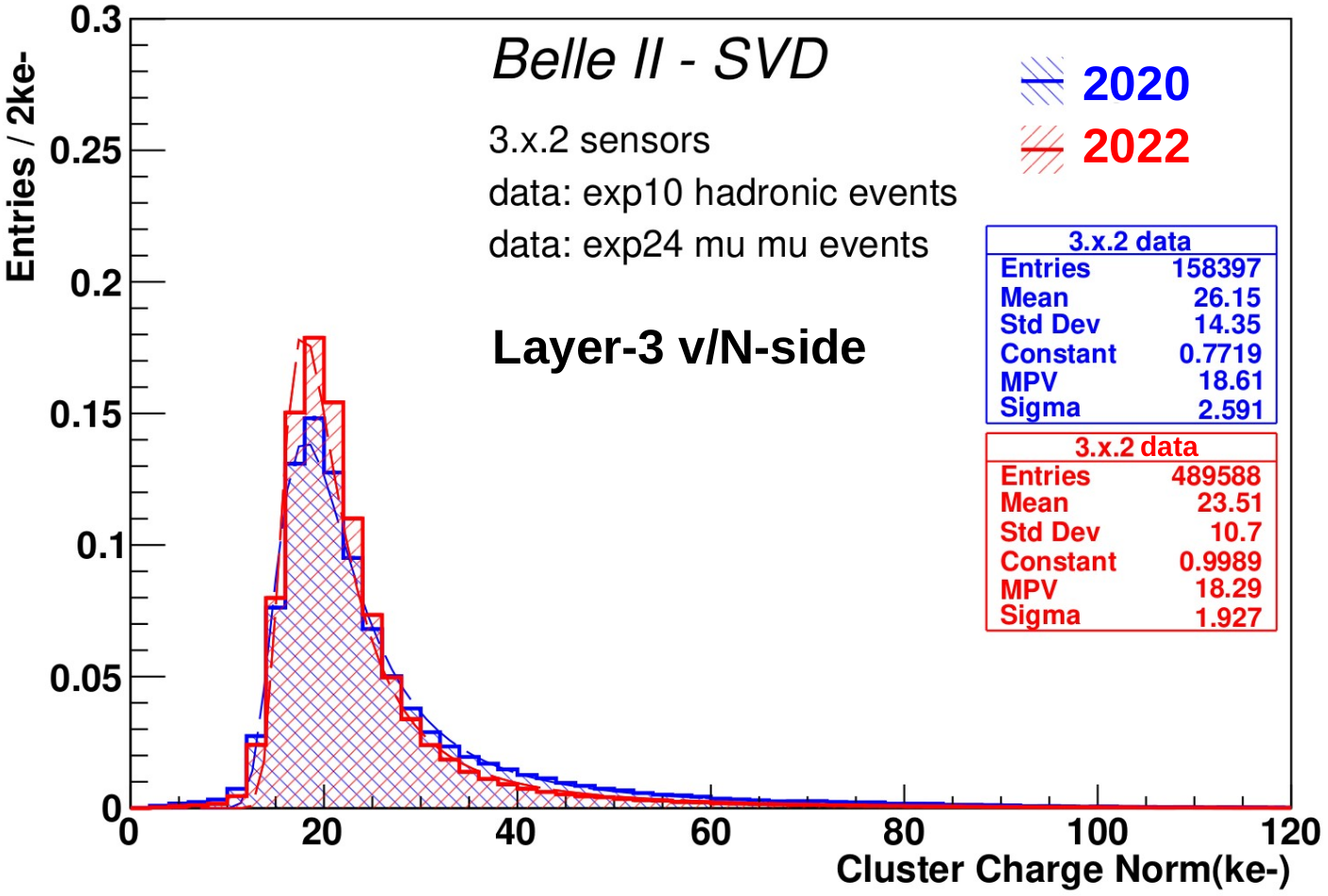}
    \caption{Cluster charge distribution for layer-3 u/P-side (left) v/N-side (right). Data collected in 2020 and 2022 are shown in blue and red respectively.} 
    \label{fig:cluster-charge}
\end{figure}

Figure\,\ref{fig:cluster-charge} shows the distribution of the cluster charge released
in the sensors in layer 3 u/P-side and v/N-side, obtained with 2020 and 2022 data. 
The release of charge depends on the track incident angle and thus it is normalised
with the track length and scaled to the sensor thickness. The fit with the Landau distribution
returns a most probable value (MPV) of \SI{21}{ke^-}. Accounting for {15}{\%}  uncertainty due to the calibration of the  gain, it is in agreement with \SI{24}{ke^-} expected for a minimum ionising particle (MIP) signal in the \SI{320}{\um} thick sensor.

\begin{figure}[!htb] 
  \centering
    \includegraphics[width=0.45\linewidth]{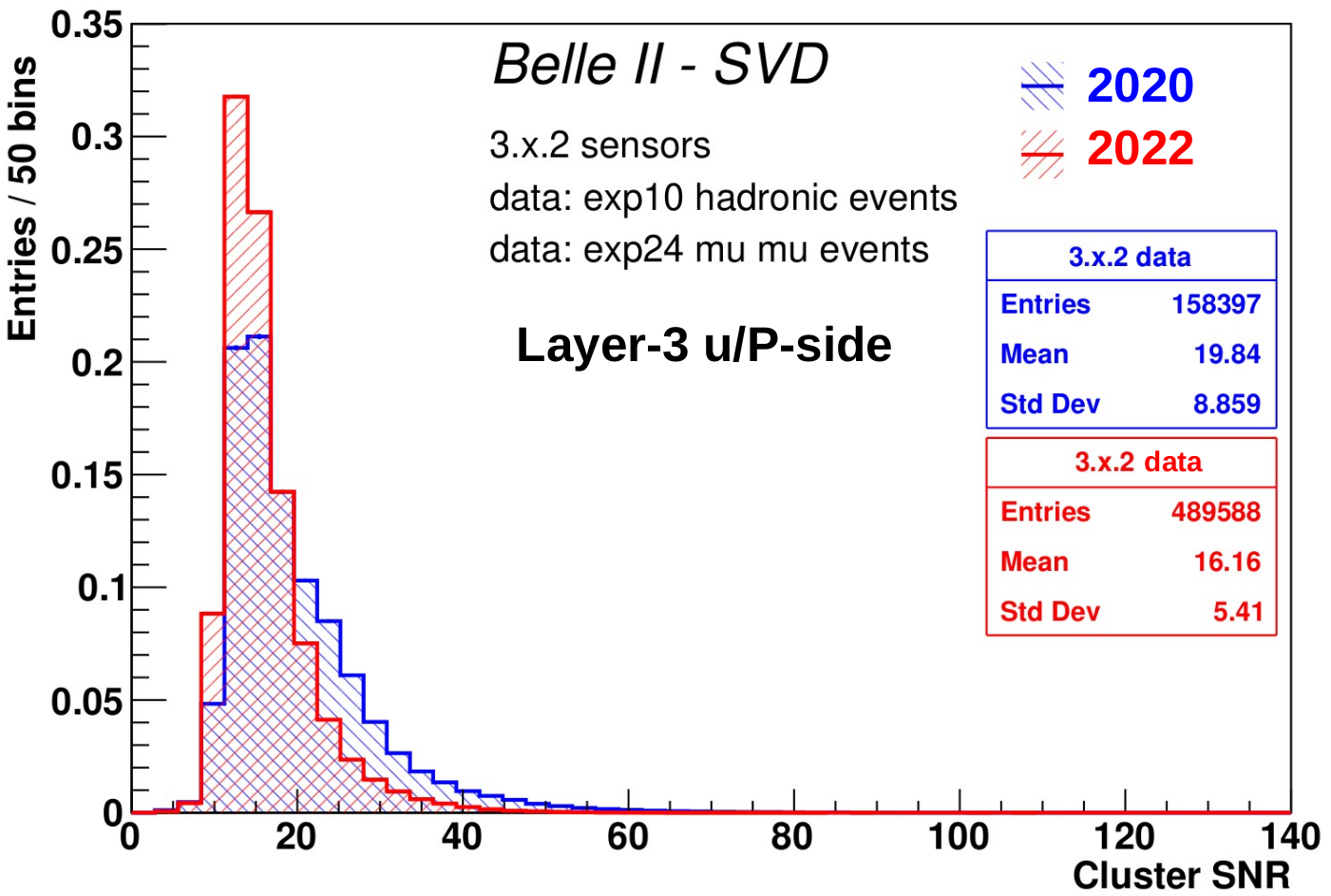}
    \includegraphics[width=0.45\linewidth]{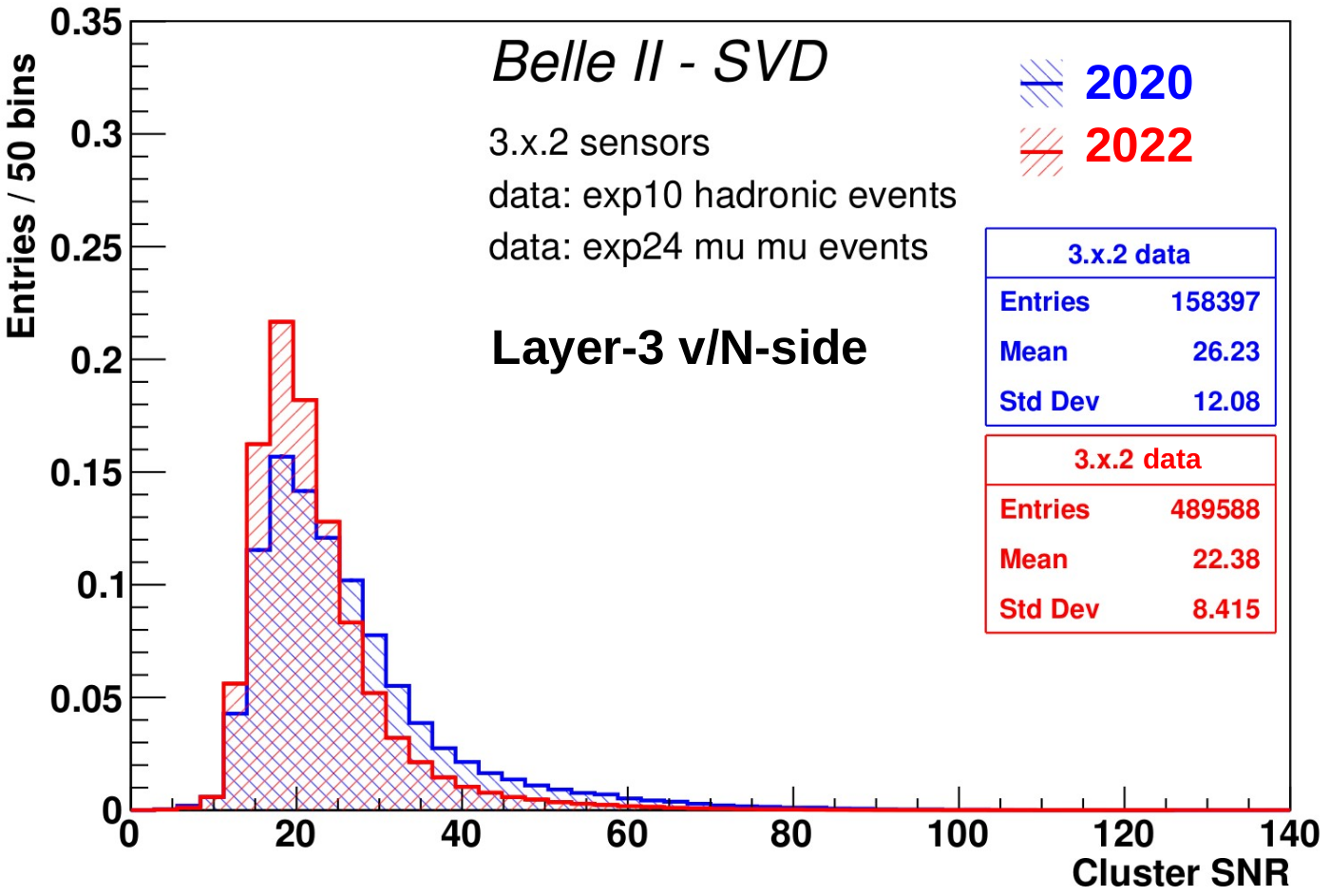}
    \caption{Cluster SNR distribution for layer-3 u/P-side (left) v/N-side (right). Data collected in 2020 and 2022 are shown in blue and red respectively.} 
    \label{fig:cluster-snr}
\end{figure}

The cluster signal-to-noise-ratio (SNR) is  
defined as the total charge collected in the cluster divided by the squared-root of the
quadratic-sum of the noises in the strips associated with that cluster.  
The cluster SNR distributions, measured in 2020 and 2022 data,
for the sensors in layer 3 u/P-side and v/N-side, are shown in Figure\,\ref{fig:cluster-snr}.
All the DSSDs perform excellently in terms of cluster SNR with MPV between 13-30 depending on
the sensor position, due to the different track incident angle, and sides, with u/P side longer strips that have higher noise.
A small reduction in the cluster SNR has been observed in the 2022 data due to increased noise from the radiation damage, at the expected level. The SNR distribution is also slightly wider in 2020 data due to the different physics events used: tracks from hadronic events (used in 2020) span a wider range for the track incident angle with the sensors, and thus have a wider cluster charge distribution, than the tracks from di-muon events used in 2022.

\begin{figure}[!htb] 
  \centering
    \includegraphics[width=0.45\linewidth]{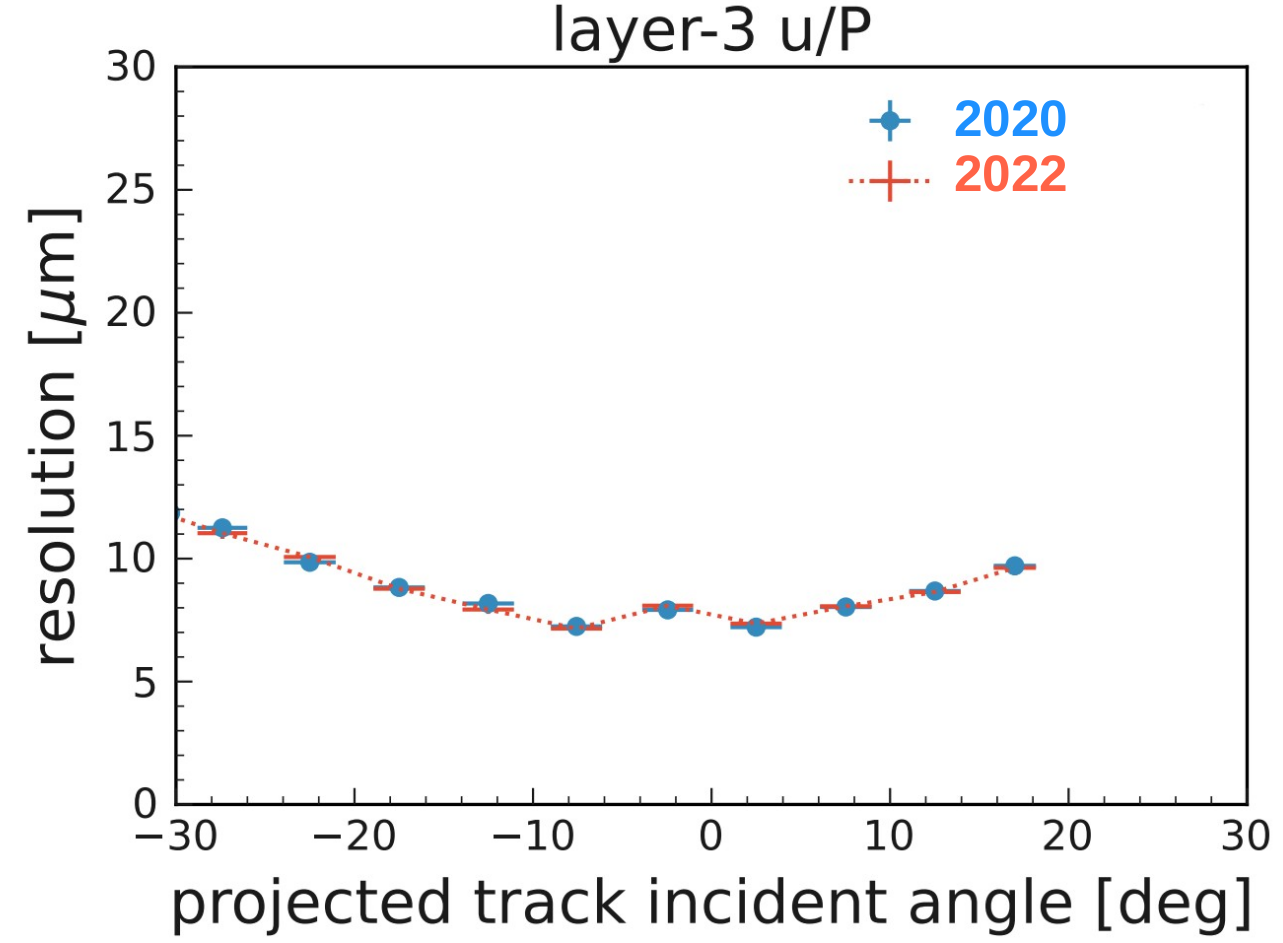}
    \includegraphics[width=0.45\linewidth]{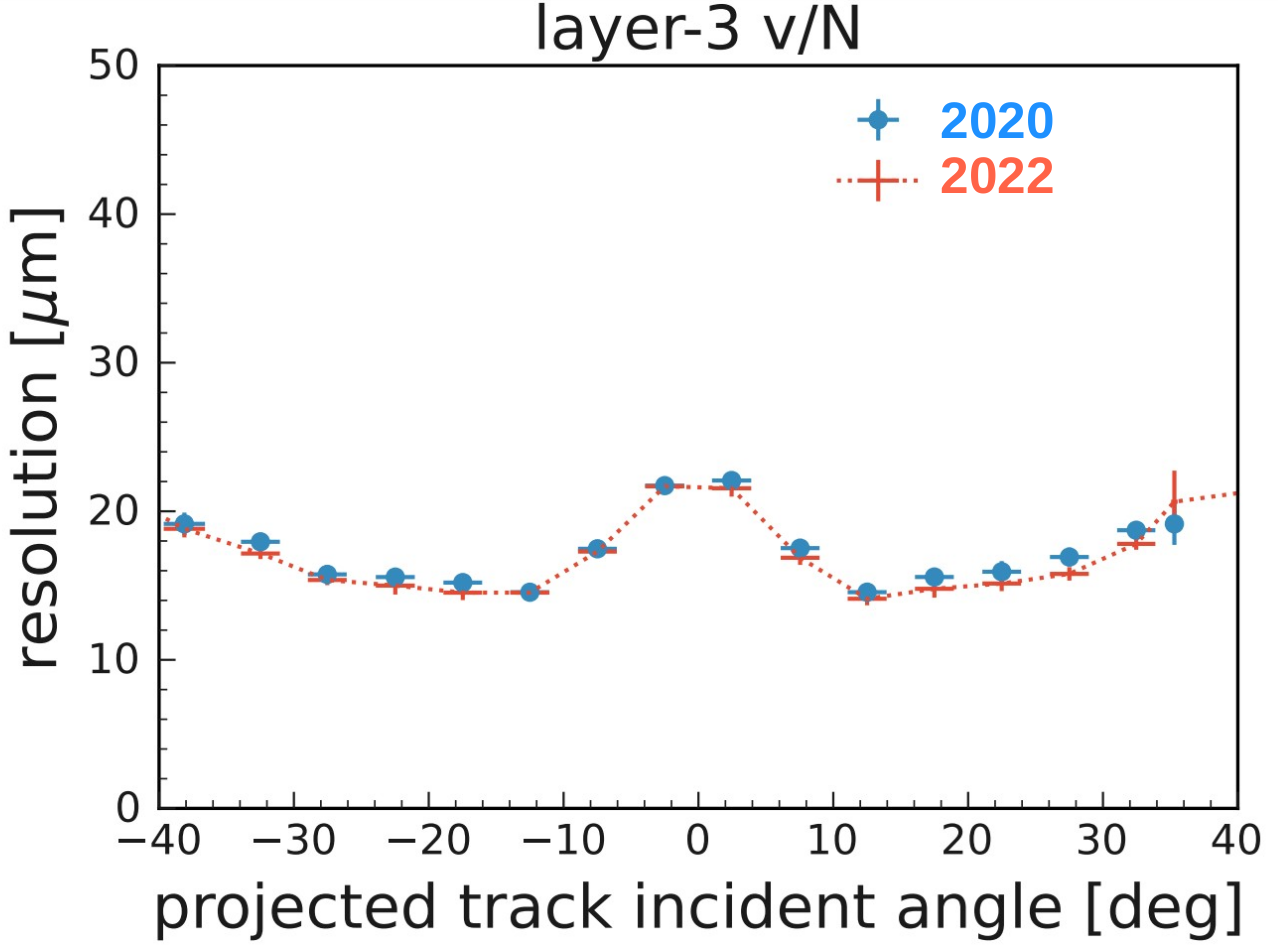}
    \caption{Cluster position resolution as a function of the projected track incident angle for layer-3 u/P-side (left) v/N-side (right). Data collected in 2020 and 2022 are shown in blue and red respectively.} 
    \label{fig:cluster-position}
\end{figure}

The cluster position resolution is estimated from the residual of cluster position against
unbiased track extrapolation using $e^-e^+ \rightarrow \mu^-\mu^+$ events. The resolutions are 
7-\SI{12}{\um} for u/P strips and 15-\SI{25}{\um} for v/N strips, which are in good agreement with
expectations. The cluster position resolutions for layer-3 u/P-side (left) and v/N-side (right) are shown in  Figure~\ref{fig:cluster-position}. Both results obtained with 2020 and 2022 data are shown.  
The SVD is constantly monitored, performance is stable and the cluster position resolution, charge and SNR are consistent over its operation period.

The SVD offers an excellent hit time measurement, explained in detail in \cite{adamczyk2022design}.
The hit time resolution is
measured from the residuals of the hit time with respect to the time of the $e^+e^-$ collision ({\it EventT0})  calculated by
other detectors with very good time resolution ($\leq \SI{1}{\ns}$). The measured hit time resolution is \SI{2.9}{\ns}
(\SI{2.4}{\ns}) for the u/P (v/N) side.

\section{Exploiting the Hit Time to Improve Reconstruction}

\begin{figure}
  \centering
  \includegraphics[width=0.6\linewidth]{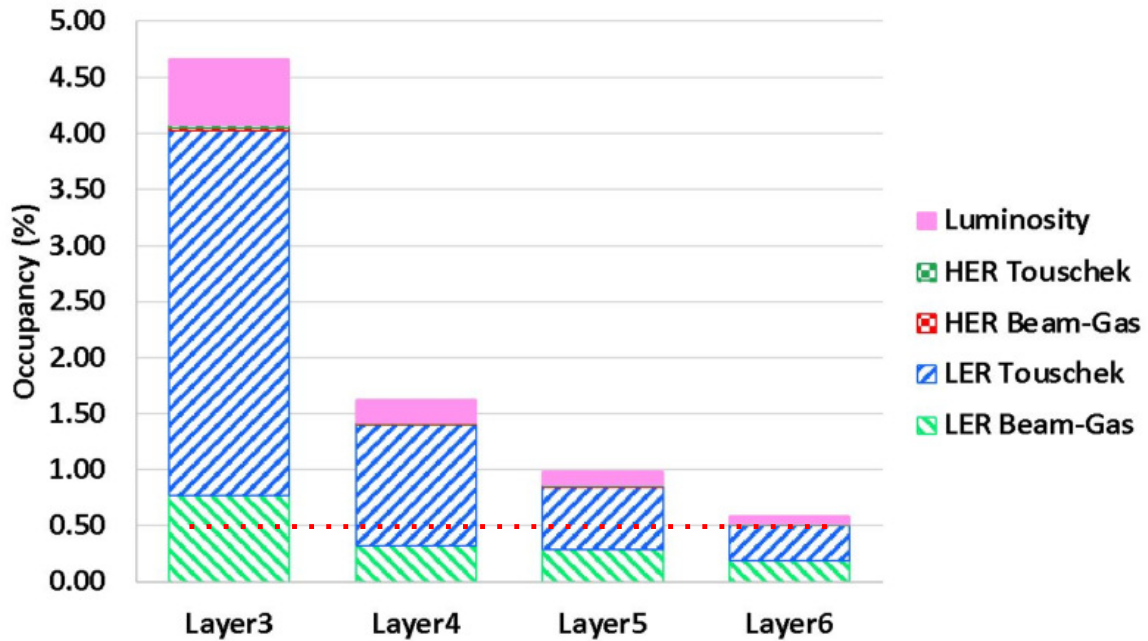}
  \caption{Expected SVD hit-occupancy on layer 3 at nominal luminosity (\SI{6e35}{\centi\meter^{-2}\second^{-1}}). }
  \label{fig:svd-occupancy}
\end{figure}

\begin{figure}
  \centering
  \includegraphics[width=0.5\linewidth]{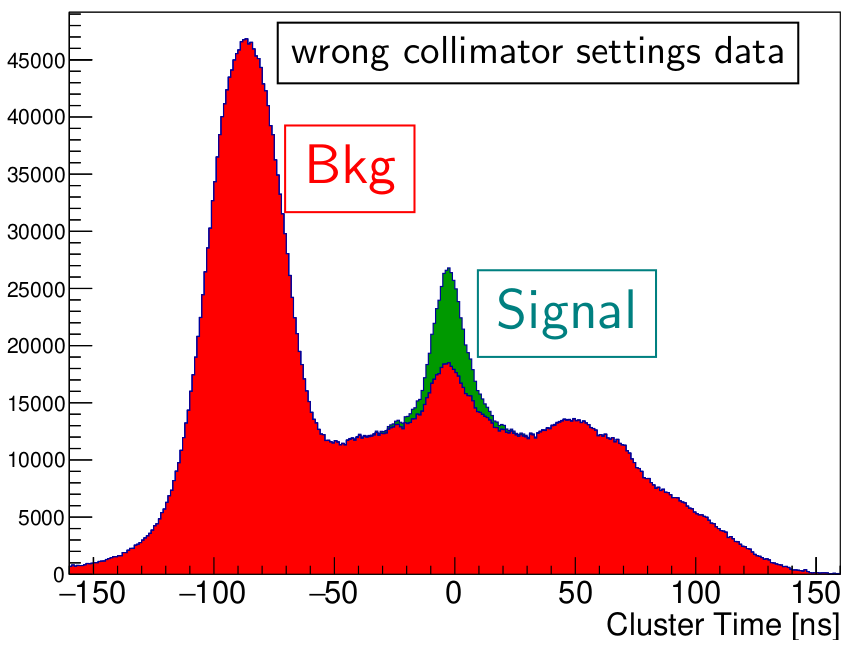}
  \caption{SVD hit time distributions (stacked) in a high background run.}
  \label{fig:hit-time-distro}
\end{figure}

One of the main challenges of the high luminosity is handling the large detector occupancy that affects the reconstruction performance. 
In the last period of data taking, the average hit occupancy in layer 3 was
less than {0.5}{\%}: however, it is expected to be almost one order of magnitude higher ({4.7}{\%}) at the nominal luminosity due to
the increased beam-induced background. Figure~\ref{fig:svd-occupancy} illustrates the expected occupancy at each SVD layer under the nominal extrapolation,  4.7\% in layer 3. However, background simulations are affected by large uncertainties related to a possible machine evolution as well as a possible redesign of the interaction region. For example, a more conservative extrapolation yields a layer 3 occupancy as high as 8.7\%. 
The SVD hit time is going to play a crucial role in keeping the current excellent tracking performance:
the  hit-time resolution for signal hits is less than \SI{3}{\ns}, which is much
smaller than the time range of beam-induced background hits (around \SI{100}{\ns}) and thus it is
very useful for background suppression.
The distribution of the time of the hits from a high-background run is shown in Figure~\ref{fig:hit-time-distro}\footnote{These data have been taken with wrong collimator settings and are affected by extremely high backgrounds; they are chosen just to show the capability of track reconstruction  when the time information is used.}. Signal hits are {\it on time}, i.e., they accumulate around the trigger time, $t=\SI{0}{\ns}$. Hits from beam-induced background, instead, are uniformly distributed in time since the bunch-crossing frequency is almost one order of magnitude higher than the APV25 readout frequency. The large bump at around $t=-\SI{80}{\ns}$ and the tail on the right of the integration window reflect the asymmetric shape of the APV25 waveform: the former is due to the accumulation of clusters created by background particles hitting the detector before the start of the acquisition, while the latter is due to missed hits outside the acquisition window.

The hit time is already employed in the SVD-only pattern recognition algorithm~\cite{Bertacchi:2020eez}. In particular, the differences of cluster times are used in filters trained on simulation and applied when creating pairs or triplets of clusters during the track candidate reconstruction. 
These filters greatly reduce the random combinations of clusters when forming a track, and help avoid using background hits for signal tracks. The filters, though, can't reject  {\it off-time} real track candidates from beam-backgrounds since their clusters have compatible times. Moreover, at high-luminosity, the combinatorics increases significantly because of the higher detector occupancy, and also the number of {\it off-time} real tracks increases, consequently these filters are not enough to assure the tracking performance needed for physics. To reduce the load on the filters, certain pre-selection criteria have been developed.

\begin{figure}
  \centering
  \begin{subfigure}[t]{0.49\textwidth}
    \includegraphics[width=1.0\linewidth]{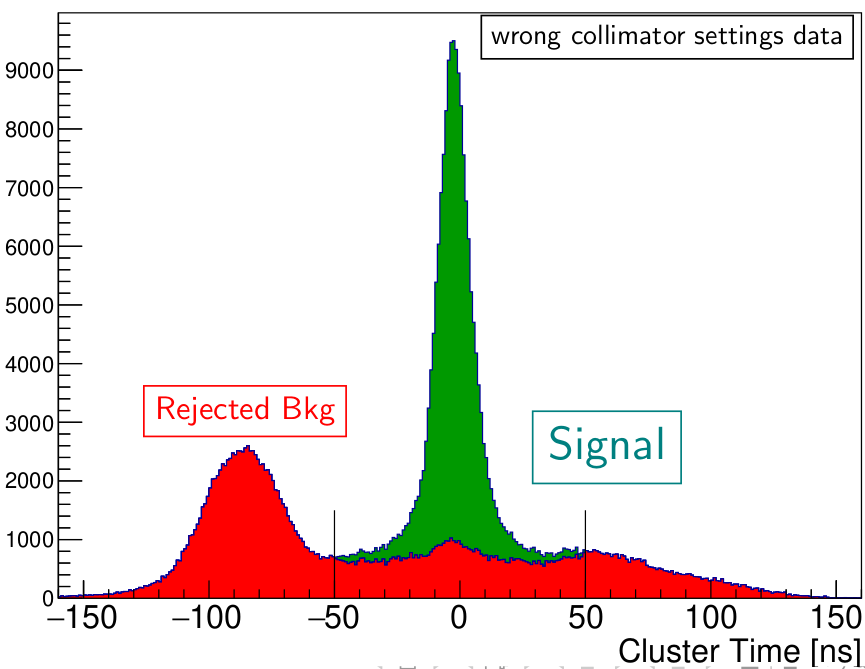}
    \caption{background rejection using {\it HitTimeSelection}.}
    \label{fig:hit-time-distro-OnOn}
  \end{subfigure}
  \begin{subfigure}[t]{0.49\textwidth}
    \includegraphics[width=1.0\linewidth]{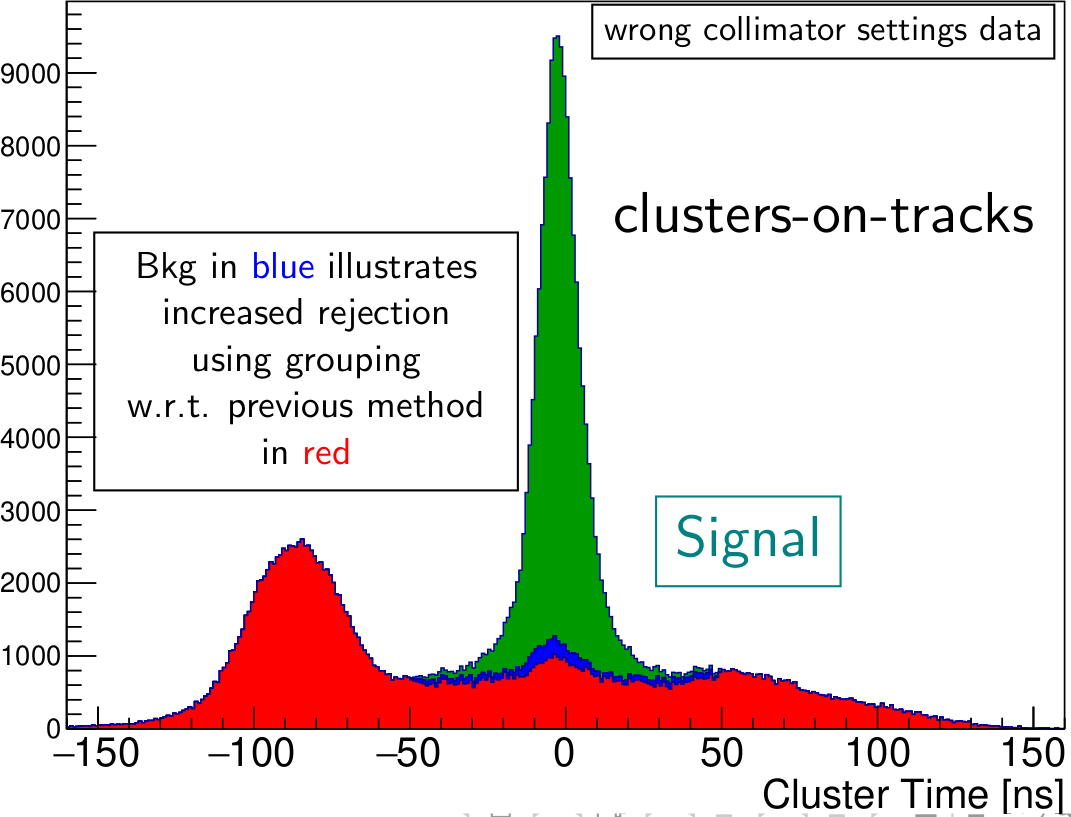}
    \caption{background rejection using {\it ClusterGrouping}.}
    \label{fig:hit-time-distro-Grouping}
  \end{subfigure}
  \caption{ SVD hit time distributions (stacked) in a high background run (the same as Figure~\ref{fig:hit-time-distro}) after the application of two different pre-selections on clusters before the tracking.}
  \label{fig:hit-time-distro-rejection}
\end{figure}

The {\it HitTimeSelection} is applied when pairing clusters from the two sides of the sensors, and it uses two simple time cuts:
$\left|t_{u,v}\right|<\text{\SI{50}{ns}}$ and $\left|t_{u}-t_{v}\right|<\text{\SI{20}{ns}}$,
where $t_{u (v)}$ is the time of u/P (v/N) side cluster. With this selection, it is possible to
suppress 50\% of background hits, while retaining 99\% of hit efficiency.
Figure~\ref{fig:hit-time-distro-OnOn} compared to Figure~\ref{fig:hit-time-distro} shows the effect of this selection on data, except for the $\left|t_{u,v}\right|<\text{\SI{50}{ns}}$ cut, indicated by vertical lines on the plot. The {\it HitTimeSelection} allows setting the layer 3 occupancy limit for good tracking to around {4.5}{\%}, which is almost the same as the expected one at the target luminosity with the nominal extrapolation (shown in Figure~\ref{fig:svd-occupancy}), leaving no safety margin to spare.

\begin{figure}
  \centering
  \includegraphics[width=0.6\linewidth]{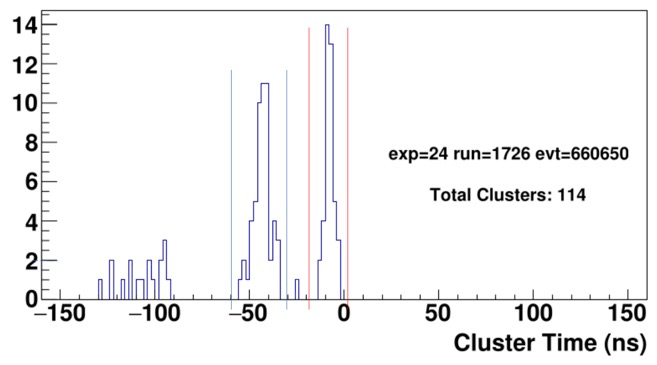}
  \caption{Cluster time distribution in one single event taken in 2022, signal (enclosed by the red vertical lines) and background groups of clusters are clearly visible.}
  \label{fig:full-event-illustration}
\end{figure}

An alternative pre-selection, namely the {\it ClusterGrouping}, has been recently developed and it is under test. This method better exploits the information provided by the cluster times, classifying the
clusters in groups on an event-by-event basis. An example is shown in
 Figure~\ref{fig:full-event-illustration} where contributions from at least two bunch-crossings can be seen.
In this method, each cluster is represented by a normalised Gaussian centered at the time of the
cluster, with a width equal to the cluster-time resolution (stored in the database). The single event is thus represented as the sum of all the Gaussian associated to the clusters in that event, with the result of {\it smoothing} the original cluster time histogram.
Each group of time-correlated clusters forms a peak in the smoothed distribution; each peak $j$ is identified and fitted with a Gaussian with a floating width $\sigma_j$. Clusters are tagged in the group $j$ if they are within $7\sigma_j$ from the $j^{\rm th}$ peak. In this procedure, clusters may belong to more than one group. Only the most populated group is finally selected for tracking, after re-weighting the group size with an exponential function of lifetime \SI{30}{\ns} (approximately \SI{3}{\times} the trigger jitter). 
Figure\,\ref{fig:hit-time-distro-Grouping} shows the background rejected using this
method.

The {\it ClusterGrouping} selection is significantly better than the {\it HitTimeSelection} as it allows further reduction of the rate of fake tracks by {16}{\%}, while keeping the same tracking efficiency. For this reason it was decided to have it in the default reconstruction in place of the {\it HitTimeSelection}.

Once the tracks are reconstructed, the SVD hits are used to compute the {\it track time} as the average time of the clusters belonging to the outgoing arm of the track, relative to the time of the collision, {\it EventT0}. 
The {\it ClusterGrouping}, together with a selection based on the track time ($|t_{\rm track}| < \SI{20}{\ns}$) that further helps to reduce the fake-track rate, allows  setting the layer 3 occupancy limit for good tracking
at around {6}{\%}, leaving some safety margin for the nominal extrapolation at the target luminosity. Further software improvements and optimizations are planned. Considering the large uncertainty on background extrapolation, and to accommodate a possible new design of the interaction region that is currently under evaluation, a future upgrade of the detector is under study~\cite{Bettarini}.

Finally, the SVD hits are  also used to compute the time of the collision, {\it EventT0}, as the average time of the clusters belonging to high transverse-momentum tracks. The resolution achieved on data is \SI{\sim1}{\ns}. This is similar to the resolution achieved with the central drift chamber (CDC), but it has a great advantage: the computation time of SVD event-time is extremely fast, about 2000 times faster than the computation of the event-time with the CDC. This feature is of particular importance as the full reconstruction chain is run also on the software-based high-level-trigger, where execution time and memory consumption have strict limits to allow processing the events incoming from the L1 trigger. For this reason it was decided to replace the CDC-based event-time computation with the SVD-based one.

\section{Activities on SVD During the Long Shutdown}

\begin{figure}
  \centering
  \includegraphics[width=1.0\linewidth]{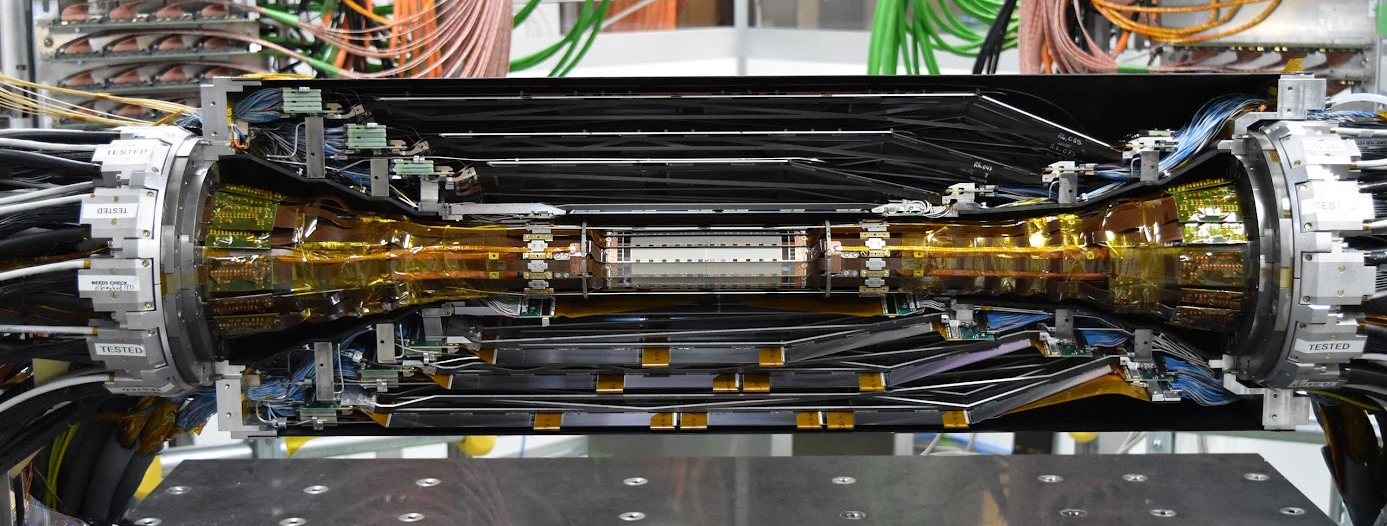}
  \caption{Photograph of the VXD after the installation of the new PXD2.}
  \label{fig:svd-half-ls1}
\end{figure}

The Belle II experiment paused its operation in July 2022 to allow for maintenance work of the accelerator and
improvements of detector, in particular the installation of a brand new pixel detector (PXD2). This period is referred to as long shutdown 1 (LS1). Intense hardware activities involved the SVD crew during the de-installation
and re-installation of VXD.
A detailed timeline can be seen in Table\,\ref{table:ls1-activity}.
The VXD was extracted from Belle~II in May 2023 and moved to the clean room to separate the half-shells
in order to access the PXD. The SVD was then tested to ensure the healthiness of all the sensors.
Once the new PXD2 was installed, the VXD half-shells were closed again. A photo of the SVD half-shell and the full PXD2 can be seen in Figure\,\ref{fig:svd-half-ls1}.
In July 2023,
the new VXD was installed in Belle~II and a commissioning phase with cosmic rays started in
September. The activity on SVD went smoothly and no significant issues were spotted during LS1. The beam operation in expected to be resumed in December 2023.

\begin{table}
  \centering
  \begin{tabular}{ccc}
  Date (2023) & Activity & Location\\
  \hline\hline
    May 10        & VXD extraction                                                                           & Belle II                                                                              \\ \hline
    May 17        & SVD detachment                                                                           & \multirow{4}{*}{\begin{tabular}[c]{@{}c@{}}activity\\ in\\ clean\\ room\end{tabular}} \\
    June 1        & SVD commissioning                                                                        &                                                                                       \\
    June 28       & New VXD assembly                                                                         &                                                                                       \\
    July 14       & New VXD commissioning                                                                    &                                                                                       \\ \hline
    July 28       & New VXD installation                                                                     & \multirow{2}{*}{Belle II}                                                             \\
    September  & \begin{tabular}[c]{@{}c@{}}Functional tests \& commissioning\\ with cosmic-ray\end{tabular} &
  \end{tabular}
  \caption{VXD re-commissioning activity during LS1 (in 2023) with the new PXD2.}
  \label{table:ls1-activity}
\end{table}

\section{Conclusion}

The Belle II SVD is operating smoothly and performing well since March 2019.
Excellent SVD performance is observed in data with greater than {99}{\%} hit efficiency on all
the sensors and less than {1}{\%} of strips masked. The cluster position resolution is very good:
7-\SI{12}{\um} for u/P strips and 15-\SI{25}{\um} for v/N strips. Performance in terms of
cluster charge and signal-to-noise-ratio is observed to be satisfactory and stable over the operation period. The effect due to radiation on the detector is observed to be as per expectation and under control.
Also the measured hit-time resolutions are very good: \SI{2.9}{\ns} (\SI{2.4}{\ns}) for
the u/P (v/N) side. Accurate and precise hit-time is crucial for the track reconstruction in high occupancy environment. The recently developed {\it ClusterGrouping} and {\it track time} allow setting the layer 3 occupancy limit at {6}{\%}, above the occupancy expected at nominal luminosity with nominal background extrapolation {4.7}{\%}. The faster SVD-based  computation of the {\it EventT0} reduces the tracking processing time on the high-level-trigger, allowing higher L1-trigger rates.

Extensive activities are performed on the VXD during the long shutdown.
A new PXD with a complete second layer is installed within the existing SVD.
The commissioning of the detector is performed with cosmic-rays, aiming to resume beam operation
in December 2023.

\section*{Acknowledgement}
\addcontentsline{toc}{section}{Acknowledgement}
This project has received funding from the European Union's Horizon 2020 research and innovation
programme under the Marie Sklodowska-Curie grant agreements No 644294, 822070 and 101026516 and
ERC grant agreement No 819127. This work is supported by MEXT, WPI and JSPS (Japan);
ARC (Australia); BMBWF (Austria); MSMT (Czechia); CNRS/IN2P3 (France); AIDA-2020 (Germany);
DAE and DST (India); INFN (Italy); NRF and RSRI (Korea); and MNiSW (Poland).

\addcontentsline{toc}{section}{References}

\bibliographystyle{jinst}
\bibliography{./IPRD23-Siena-SVD-talk-Proceeding}

\end{document}